\documentclass[journal=jpccck,manuscript=article]{achemso}
\usepackage{natbib}
\usepackage{times}
\usepackage{epsfig}
\usepackage{graphicx}
\usepackage{psfrag}
\usepackage{ae}
\usepackage{amsmath,amssymb}
\usepackage[usenames]{color}
\usepackage{float}
\usepackage{xcolor}
\usepackage{placeins}
\usepackage{amsfonts}
\usepackage{bm}
\usepackage{setspace}
\usepackage{soul}
\usepackage{soulutf8}

\large
\title{\bf Flexible model of water based on the dielectric and electromagnetic spectrum properties : TIP4P/$\epsilon \hspace{0.1cm} _{Flex}$.}
\author{Ra\'ul Fuentes-Azcatl}
\email{razcatl@correo.xoc.uam.mx}
\affiliation{ Universidad Aut\'onoma Metropolitana Unidad Xochimilco, Calzada del Hueso 1100, CDMX, CP 04960, México }

\newpage
\begin{document}

\date{}

\newpage

\maketitle
\begin{abstract}
The Infrared Spectrum is used as an experimental data target, to improved the TIP4P/$\epsilon $, adding  harmonic potential U(r) in all bonds and  harmonic potential U($\theta$) in the angle formed by the hydrogens and oxygen atoms of the water molecule.

The flexibility in the molecule gives the ability of the water molecules to change, around different temperatures and pressures, their structure and in the bulk liquid the distribution of the dipole moment. This distribution helps to reproduce better the experimental data that the rigid models can not describe. The rigid water models of  3 and 4 sites have limitations to describe all the experimental properties, and is because can not take on how the dipole moment is distributed of dipole moment around the different thermodynamics phases. \\The new flexible TIP4P/$\epsilon \hspace{0.1cm} _{Flex}$ water model is compared to the improve models of TIP4P and OPC model.
\end{abstract}

\section{Introduction}

The research of how the
 molecular interactions in water are related to its unique properties, is a fundamental and important  subject  in physical chemistry. A related essential question is how water interacts with solutes such
 as ions\protect\cite{nacle,kbre}, organic molecules \cite{edgar} and proteins \cite{Georgoulia}. These reasons are a field of study of water interactions and has been a highly active field for decades in experimental and theoretical chemistry.

Several interesting and successful interaction potentials (force fields, FF) have
been developed in the last years\cite{Guillot,Vega11}. The simulation research had been showed that these
FF take on for many important properties of liquid water,  although, is there still space to improvement the rigid models?. Now a days the TIP4P/$\epsilon $\cite{tip4pe} can reproduce better the experimental values than others \cite{marcia1,Piotr}. 

In molecular simulations, solvents such as water are often
represented by explicit models such as TIP3P or SPC/E.\cite{ tip4p, Berendsen} Many of these explicit-solvent models were originally developed and parametrize a few decades ago. They were parametrize to
agree with experimental data such as the density, self-diffusion coefficient and enthalpy of vaporization of liquid phase, key quantities that are readily
available from accurate experiments. However, such molecular solvent models do not typically predict the correct experimental values of transport and  dielectric properties.\cite{Vega11, marcia1,Piotr}
This is unfortunate because
molecular simulations are so commonly used to treat the solvation of polar and charged molecules, for which it is
essential that the solvent respond properly to electrostatic fields.

Since the TIP4P model\cite{tip4p} was improved, by adjusting the epsilon parameter of the Lenard-Jones potential to obtain an important improvement by the experimental properties able to reproduce, TIP4P/05 model\cite{vega05}. The search has continued to have a better model that reproduces more and better the experimental values at different pressure and temperature conditions.

By applying the dipole moment of minimum density  method \textbf{$\mu_{min\rho}$} to the original TIP4P model, it was improved and was name TIP4P/$\epsilon$\cite{tip4pe,marcia1}(because reproduce the dielectric constant at different thermodynamics conditions), as well as the SPC model was improved with the same method and the SPC/$\epsilon$\cite{spce} was obtained.
TIP4P/$\epsilon$ model can reproduce the same experimental data as the TIP4P/05 and adds the reproduction of the experimentla values of the dielectric constant with a minimal of error\cite{Piotr} and isothermal compressibility under various thermodynamic conditions. It seems that this model is the most optimal under this rigid 4-site scheme, as shown by the work of Wang et al\cite{ping}, reaching very close parameters to the TIP4P/$\epsilon$ with computational machine learning methods. Another effort to improve the TIP4P force field is the recent TIP4P-ST by Qiu et al\cite{tip4pst}; TIP4P-ST implement the method of Salas et al\cite{salas}, using the surface tension as a target parameter. In this work these last two models are compared with the new TIP4P/$\epsilon \hspace{0.1cm}_{Flex}$

A common feature of the most popular water models is that they are rigid non-polarizables, i.e., the intramolecular degrees of freedom are off.

It may seem obvious that a step forward for the improvement of the rigid non-polarizable water potentials would
be the addition of flexibility. However, there has been some skepticism in the past about the usefulness of flexible water models.\cite{Smith, Tironi, Gonzalez, Kurisaki} In fact, Tironi et al.\cite{Tironi} concluded that the introduction of flexibility creates more problems than it solves and  does not improve upon the accuracy of rigid models. 
It seems natural that the development of flexible models  will consist of the addition of flexibility to a successful rigid potential.\\

A number of computer simulations with different flexible
water models have been reported \cite{Marti, Wu, Teleman, Barrat, Guillot, Wallqvist, Lemberg, abascal} since the pioneering
works of Lemberg and Stillinger \cite{Lemberg} and Toukan and Rahman.\cite{Toukan}

The dominant paradigm in water model development is to fit the parameters to reproduce a set of experimentally measured condensed phase properties. Generally
speaking, a diverse data set over a wide range of thermodynamic conditions improve the
domain of applicability of the model, but it also increases the complexity of the optimization problem. \\
The last proposal is the one made by Fuentes and Barbosa FAB/$\epsilon$\cite{fabe}, taking into account the flexibility through the addition of a harmonic potential in the OH bond and another in the HOH angle of a 3-body model, generating an improvement in the models of three sites. By adding the harmonic potential in the bond of a molecule such as CO2, it has been possible to improve the reproduction of experimental values\cite{co2e}

The rest of the work is organized as follows: Section 2 gives the force field of TIP4P/$\epsilon \hspace{0.1cm} _{Flex}$ model of water, Section 3 search of parameters 4 gives the Results and in Section 5 the conclusions. Finally, references are given.\\

\section{The force field of TIP4P/$\epsilon \hspace{0.2cm} _{Flex}$ model of water}

As for the intermolecular potential, we used a four sites model like the usual
choice of TIP4P model structures: a Lennard-Jones center at the
position of the oxygen atom plus the electrostatic interaction
given by two positive charges located at the hydrogen atoms
and a compensating negative charge placed at the so-called
M-site. 
The water molecules have four sites: two hydrogens, an oxygen and a site M located at a distance $l_{OM}$ from the oxygen atom along the bisector of the hydrogen atoms. The intramolecular interactions in the flexible model are defined by harmonic potentials in bonds and angle,
 
 \begin{equation}
\label{eqn1}
U(r)\!=\! \frac{k_r}{2}(r-r_0 )^2 \quad\hbox{and}
\quad U(\theta)\!=\!\frac{k_{\theta}}{2}(\theta-\theta_0)^2 ,
\end{equation}

\noindent where $r$ is the bond distance and $\theta$ is the bond angle. The subscript $0$ denotes their equilibrium values, $k_r$ and $k_{\theta}$ are the corresponding spring constants. \\
The intermolecular force field between two water molecules is based on the LJ and Coulomb interactions,

\begin{equation}
\label{ff}
u(r) = 4\epsilon_{\alpha \beta} 
\left[\left(\frac {\sigma_{\alpha \beta}}{r}\right)^{12}-\left (\frac{\sigma_{\alpha \beta}}{r}\right)^6\right] + \frac{1}{4\pi\epsilon_0}\frac{q_{\alpha} q_{\beta}}{r}
\end{equation}

\noindent where $r$ is the distance between sites $\alpha$ and $\beta$, $q_\alpha$ is the electric charge of site $\alpha$, $\epsilon_0$ is the permitivity of vacuum,  $\epsilon_{\alpha \beta}$ is the LJ energy scale and  $\sigma_{\alpha \beta}$ the repulsive diameter for an $\alpha \beta$ pair. The cross interactions are obtained using the Lorentz-Berthelot mixing rules,

\begin{equation}
\label{lb}
\sigma_{\alpha\beta}= \left(\frac{\sigma_{\alpha\alpha} + \sigma_{\beta\beta} }{2}\right);\hspace{1.0cm} \epsilon_{\alpha\beta}= \left(\epsilon_{\alpha\alpha} \epsilon_{\beta\beta}\right)^{1/2}
\end{equation}

The position of site M is ${\bf r}_M={\bf r}_O + a({\bf r}_{H_1}-{\bf r}_O)+a({\bf r}_{H_2}-{\bf r}_O)$ where $a=l_{OM}/[2r_{OH} cos(\theta/2)]$ for rigid molecules. The force of site M is distributed among the other atoms according to: ${\bf F} (H_1)=a{\bf F}(M)$, ${\bf F}(O)=(1-2a){\bf F} (M)$ and ${\bf F}(H_2)=a{\bf F}(M)$.  In the flexible model the average of the $r_{OH}$ distance is used to calculate $l_{OM}$.\\

The parameters used in the water models
given in table \ref{table1}. The calculus  does not consider the nuclear quantum effects and intermolecular zero-point energy quantum fluctuations\cite{Ceriotti,Paesani}. It is known that the above affects the structure and dynamics of the hydrogen-bonding structure in liquid water via tunneling and proton delocalization.

\begin{table}
\caption{ }
\label{table1}
\scalebox{0.9}[1]{
\begin{tabular}{|lccccccccc|}
\hline\hline
model	&	$k_{b}$	&	$r _{OH_{eq}}$	&	 $k_{a}$	&	    $\Theta_{eq}$ 	&	d$_{OM}^{rel}$	&	$\varepsilon_{OO}$ 	&	$\sigma_{OO}$  	&	$q_{O}$	&	$q_{H}$ 	\\
	&	kJ/ $mol$ {\AA}$^{2}$ 	&	{\AA}	&	kJ/ $mol$ rad$^{2}$	&	deg	&		&	/kB	&	{\AA}	&	e	&	e	\\
	\hline\hline
FBA/$\epsilon $	&	3000	&	1.0270	&	383	&	114.70	&	-	&	95.29998	&	3.17760	&	-0.845	&	0.42250	\\
TIP4P/$\epsilon$	&	-	&	0.9572	&	-	&	104.52	&	0.1050	&	93.00000	&	3.16500	&	0	&	0.52700	\\
TIP4P/$\epsilon \hspace{0.1cm} _{Flex}$	&	1570	&	0.9300	&	212	&	111.50	&	0.0830	&	95.50000	&	3.17340	&	0	&	0.51000	\\
OPC	&	-	&	0.8724	&	-	&	103.60	&	0.1594	&	107.09165	&	3.16655	&	0	&	0.67910	\\
TIP4P-FB	&	-	&	0.9572	&	-	&	104.52	&	0.1052	&	90.12268	&	3.16555	&	0	&	0.52587	\\
TIP4P-ST	&	-	&	0.9572	&	-	&	104.52	&	0.0989	&	89.04258	&	3.16610	&	0	&	0.52172	\\

\hline
\end{tabular}}
Notice that r$_{OH_{eq}}$ and $\Theta_{eq}$ define the rigid geometry of non-polarizable force fields\\
The charge on site M is q$_M$ = -2q$_H$  \\

\end{table}

\section{Search of parameters}

Using the dipole moment of minimum density method (\textbf{$\mu_{min\rho}$}), \cite{tip4pe} obtains the best geometry that be able to reproduce properties described in table \ref{table3}. \\
To acquire the best parameters using the \textbf{$\mu_{min\rho}$} method, in the case of the TIP4P/$\epsilon \hspace{0.1cm} _{Flex}$ is linked to obtaining the values of the harmonic potential constants, which reproduce values close to those reported experimentally for the IR spectrum.\\
The value of the harmonic potential constants (bond and angle) are referred to the reproduction of the experimental values of OH stretching and HOH bending mainly, the parameters calculated for flexible water models given in table \ref{table2}.

\begin{table}
\caption{Wavenumbers (in cm $^{-1}$ ) at the peak of the bands of the power spectrum for the flexible models at liquid phase. }
\label{table2}
\begin{tabular}{|lccccc|}
\hline\hline
	&	SPC/Fw\protect\cite{Wu,fabe}	&	FAB/$\epsilon$\protect\cite{fabe}	&	TIP4P/$\epsilon \hspace{0.1cm} _{Flex}$	&	TIP4P/05f\protect\cite{abascal}	&	Exp.\protect\cite{abascal}	\\\hline\hline
Cage  vibrations	&	50	&	50	&	50	&	50	&	50	\\
Intermolecular stretch	&	278	&	278	&	263	&	230	&	183.4	\\
Librations  A2, B2	&	513	&	542	&	548	&	570	&	430, 650	\\
Bending  (H–O–H)	&	1500	&	1600	&	1324	&	1670	&	1643.5	\\
Stretching  (O–H)	&	3685	&	3000	&	2200	&	3370	&	3404	\\

\hline
\end{tabular}
\end{table}

\begin{figure}
\centering
\centerline{\psfig{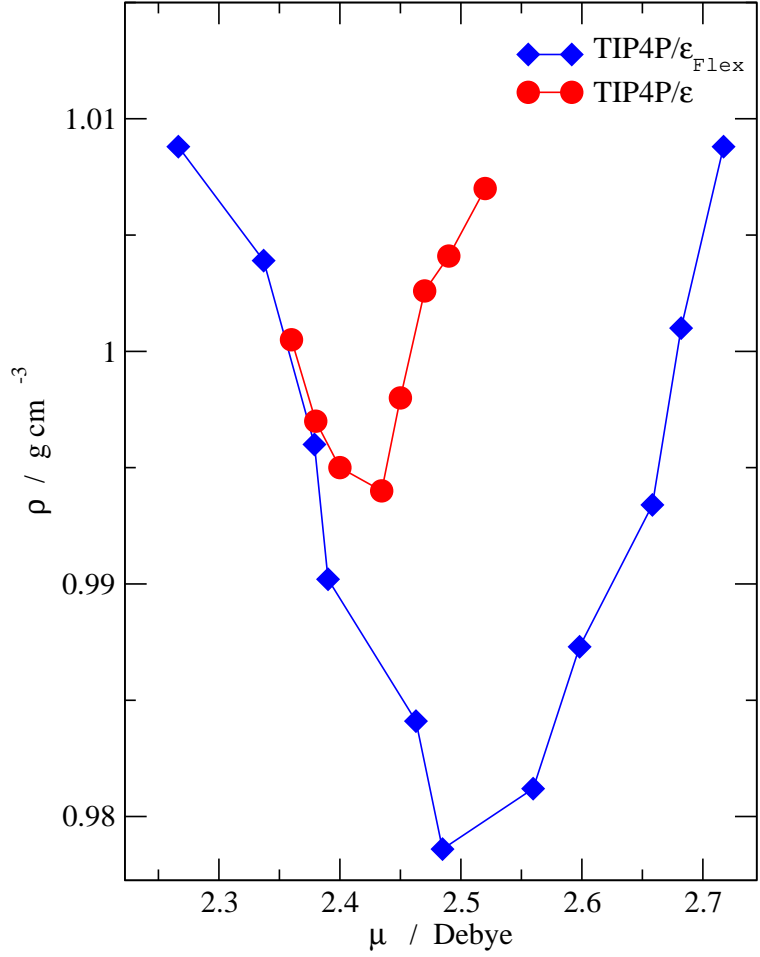}}
\caption{ Dipole moment of minimun density \textbf{$\mu_{min\rho}$} versus density for TIP4P/$\epsilon\hspace{0.1cm} _{Flex}$(blue diamonds) and TIP4P/$\epsilon$ (red circles).}
\label{MDmin}	
\end{figure}

The location of the infrared spectrum of the OH stretching is 27\% off and the location due to the HOH bending is 19\% off. No greater precision could be achieved because when applying the \textbf{$\mu_{min\rho}$} method with greater accuracy in this location, it is not possible to have an optimal \textbf{$\mu_{min\rho}$} value.\\
The dipole moment of minimum density that could be found with the aforementioned level of error and  improves the rigid non-polarizable model (TIP4P/$\epsilon$) is presented in the figure \ref{MDmin}.  The \textbf{$\mu_{min\rho}$} of TIP4P/$\epsilon \hspace{0.1cm} _{Flex}$ is located at 0.9786 g $cm^{-3}$, which is located below that obtained by the TIP4P/$\epsilon$. This gives the final data of the parameterization and the calculations are continued under various thermodynamic conditions to verify the correct parameterization.

Even though the TIP4P/05f model reproduces the correct IR spectrum, it does not compare in this work with the new model and the non-polarizable ones, because it reproduces them with less accuracy than the TIP4P/05 model\cite{abascal}.
\newpage

\newpage

\subsection{Simulation Details}

All results of all models here presented,  have been calculated with the same level of theory, parameters, and programs in the molecular dynamics simulations.

In this work have been performed molecular dynamics simulations in the isothermal-isobaric ensemble, NPT, with isotropic fluctuations of volume, to compute the liquid properties at different temperatures and pressures, $1\;bar$; these simulations involved typically 500 molecules.

To compute the surface tension has been used the
constant volume and temperature ensemble,
NVT, and   5832 molecules. To obtain the liquid-vapor interface by setting up a liquid slab surrounded by vacuum
in a simulation box with periodic boundary conditions in the three spatial directions. The dimensions of the simulation cell were $Lx = Ly = 46.68$ {\AA} with $Lz = 3Lx$, with $z$ being the normal direction to the liquid-vapor interface.
The GROMACS 2016 package~\cite{Hess, Spoel} was employed in all simulations presented in this work. The equations of motion were solved using the leapfrog algorithm with a time step of $1\;fs$ for flexible models and $2\;fs$ for non-polarizable rigid models. The temperature was coupled to the Nos\'e-Hoover thermostat with a parameter ${\tau_T}= 0.6\;ps$ while the pressure was coupled to the Parrinello-Rahman barostat~\cite{Parrinello} with a
coupling parameter ${\tau_P}= 0.5\;ps$. 

To compute the  electrostatic interactions
 with the particle mesh Ewald approach~\cite{Essmann}  with a grid spacing of $1.2$ {\AA} and spline interpolation of order 4. In the
isotropic NPT simulations, the real part of the Ewald
summation and the LJ interactions were truncated at $9$ {\AA}. Long range corrections for the LJ energy and pressure were included. The dielectric constant is obtained from the analysis of 
the dipole moment fluctuations of the simulation 
system~\cite{Neumann, Hansen}. The density ($\rho$), self-diffusion coefficient (D), isothermal compressibility 
($\kappa$), enthalpy of vaporization $\Delta H_{vap}$, thermal expansion coefficient($\alpha$), specific heat capacity C$_P$, and the dielectric constant ($\epsilon$) were calculated from the
same simulation for at least $100\;ns$ after an equilibration period of $10\;ns$. For the surface tension computations in the NVT ensemble, the cutoff was set to $26$ {\AA}, since the surface tension
depends on the truncation of the interactions \cite{Truckymchuk} and the
interface cross-sectional area.\cite{Orea, Minerva} The equilibration 
period for the interfacial simulations was $2\; ns$, and the results for the
average properties were obtained over an additional $10\; ns$ trajectory. 

For the calculation of the density of the solid phases here reported (ice Ih), have been carried out isothermal-isobaric (NpT) simulations. For the initial configurations, we used the structural data obtained from diffraction experiments. The NpT simulations are performed under periodical boundary conditions at 1fs and 2fs without seeing any change in density  at 10ns.The Berendsen thermostat and barostat were used with parameters of 0.2ps and 0.5ps, respectively

The Berensend barostat was employed for the calculation of the melting temperature and of the  density of 
the ice. The use of this barostat allows the simulation box to expand or contract, and then to form ice or liquid phases. For studying the ice phase and the 
melting properties,  the temperature was fixed with a Berendsen thermostat with a relaxation time of $0.2\;ps$~\cite{Garcia}.  For the description of the coexistence between liquid and solid water, have been employed an 
orthogonal cell. This approach is consistent with the 
crystallography data of the solid phase Ih~\cite{Petrenko}. The cell size 
is $Lx =21.6${\AA}, $Ly=23.3${\AA} and $Lz=53.8${\AA} .Which gives us a 
contact area between the $Lx*Ly=503.28${\AA}${^2}$ phases.
\newpage

\section{Results}

\begin{figure}
\centering
\centerline{\psfig{figure=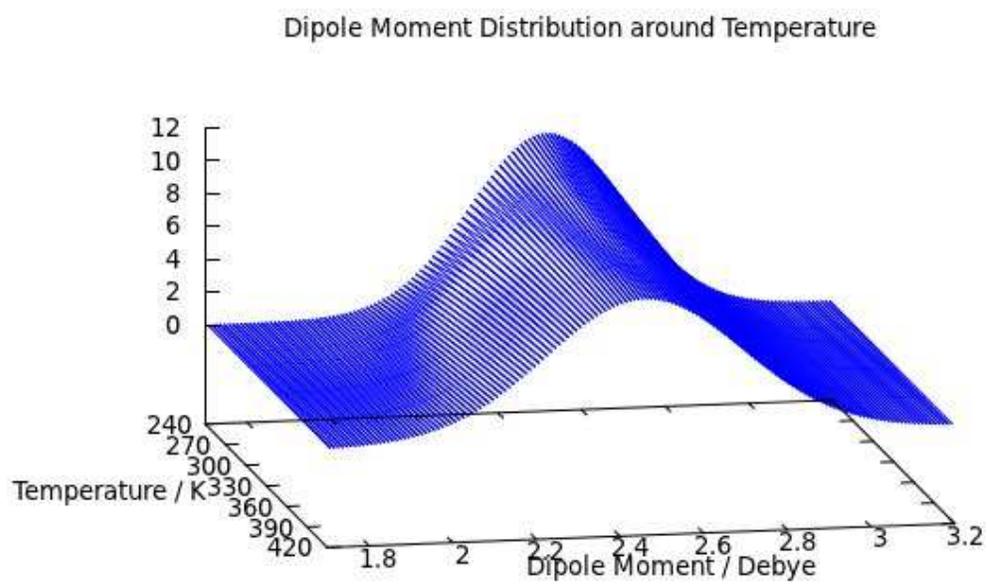,width=16.0cm,angle=0}}
\caption{Distribution of  molecular dipole moment $\mu$ versus 240K to 420K temperatures }
\label{3D-dist}	
\end{figure}
\newpage
Figure \ref{3D-dist} represent the distribution of the dipole moment DM of water calculated with the TIP4P/$\epsilon\hspace{0.1cm} _{Flex}$ in the liquid phase at different temperatures, at low temperatures the DM is closed to 2.5 debyes, the distribution is narrower and the mean value is larger; when the temperature increase the DM decrease to 2.43 debyes, the distribution is more homogeneous and the mean value is smaller

\newpage

\newpage
\begin{figure}
\centering
\centerline{\psfig{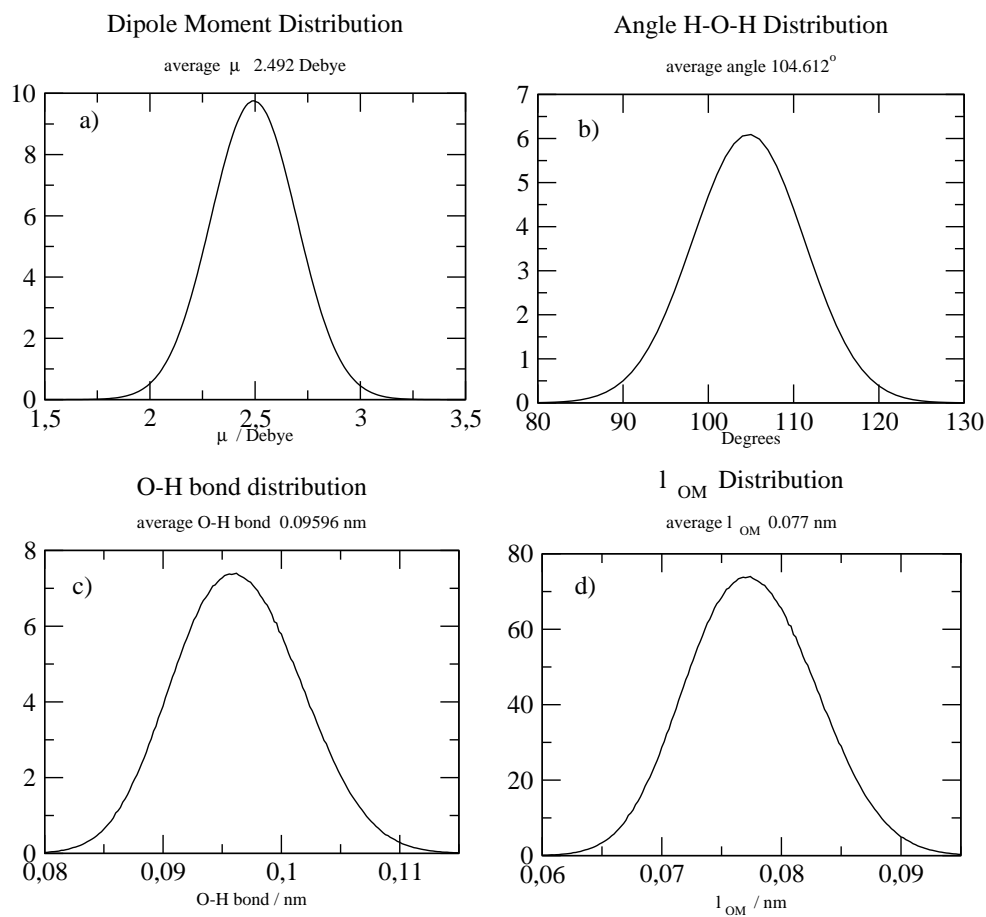}}
\caption{Distributions of flexible models of water:TIP4P/$\epsilon\hspace{0.1cm} _{Flex}$ at 298 K and 1 bar of temperature and pressure respectively, a) bond distance O-H. b) bending angle HOH. c) molecular dipole moment $\mu$. d) $l_{OM}$ distance.}
\label{distribution}
\end{figure}

\newpage

The  water  structure of TIP4P/$\epsilon\hspace{0.1cm} _{Flex}$ at 298 K and 1 bar of temperature and pressure respectively is present in figure \ref{distribution}.
 Figure~\ref{distribution}(a) shows the  distribution of dipole moments of  the water molecules with an average of 2.492 D, whereas the corresponding experimental value is the 2.42 D. 
Differently from the non-polarizable rigid models, the TIP4P/$\epsilon\hspace{0.1cm} _{Flex}$ exhibits a distribution of  HOH angles illustrated in 
 figure~\ref{distribution}(b), the average angle, $104.612^o$ is close to the average experimental value~\cite{Ichikawa} which is $106^o$.
The distribution of O-H bond distances for 
the TIP4P/$\epsilon\hspace{0.1cm} _{Flex}$ model is illustrated in the figure~\ref{distribution}(c), this result shows the average bond distance at 
 $0.09596\;nm$ what is $3\%$ lower than the  neutron diffraction value, $0.099\;nm$~\cite{Zeidler}, and
only  $2.4\%$ lower than the X-ray diffraction 
value, $ 0.09724\;nm$~\cite{Na71,To00}.
 The distribution of site M around the bisector of molecule respect to the Oxigen $l_{OM}$ for 
the TIP4P/$\epsilon\hspace{0.1cm} _{Flex}$ model is illustrated in figure~\ref{distribution}(d), this result shows the average bond distance at 
of $0.077\;nm$ 27\% less than the value of TIP4P/$\epsilon$. 
In principle, rigid models can be constructed to give this bond distance, however, they can not adapt to the thermodynamics conditions and can change with the temperature or pressure the O-H bond distance observed both in the experiments and in TIP4P/$\epsilon\hspace{0.1cm} _{Flex}$.

\newpage
\begin{figure}
\centerline{\psfig{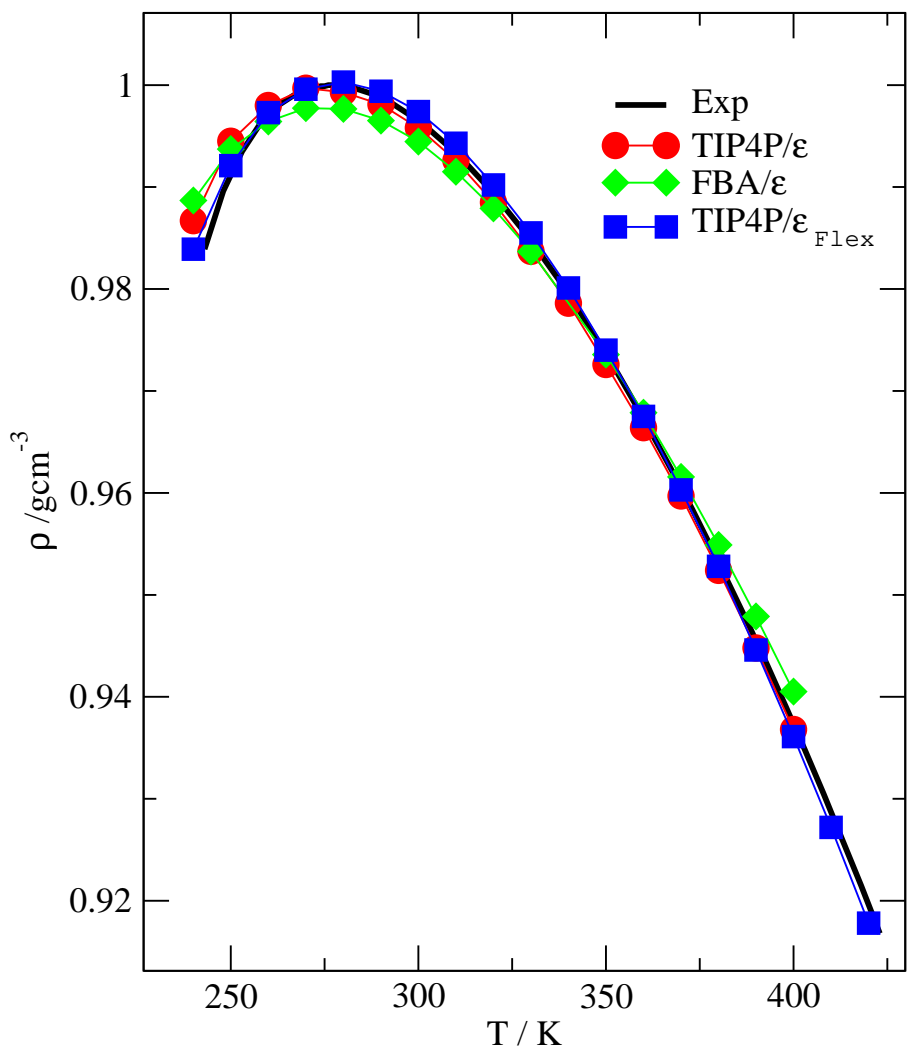}}
\caption{Density as a function of temperature at 1 bar of pressure for the TIP4P/$\epsilon\hspace{0.1cm} _{Flex}$(blue squares), FBA/$\epsilon$ (green diamonds) and TIP4P/$\epsilon$ (red circles) models and experimental data~\cite{NIST} (solid line). }
\label{rho}
\end{figure}
\newpage
The liquid densities as a function of temperature and 1 bar are shown in Figure \ref{rho} for the TIP4P/$\epsilon\hspace{0.1cm} _{Flex}$, FBA/$\epsilon$ and TIP4P/$\epsilon$ water models. All the simulation details are the same; the difference is the value of the force fields. The three models, shown in figure \ref{rho}, give the same $\rho$-T shape and around the same TMD. The experimental results for temperatures below 273 K are for the metastable liquid. Error concerning the experimental value of the TIP4P/$\epsilon\hspace{0.1cm} _{Flex}$ model is less than 0.5\% in the liquid phase, as expressed in table \ref{table3}.


\newpage
\begin{figure}
\centerline{\psfig{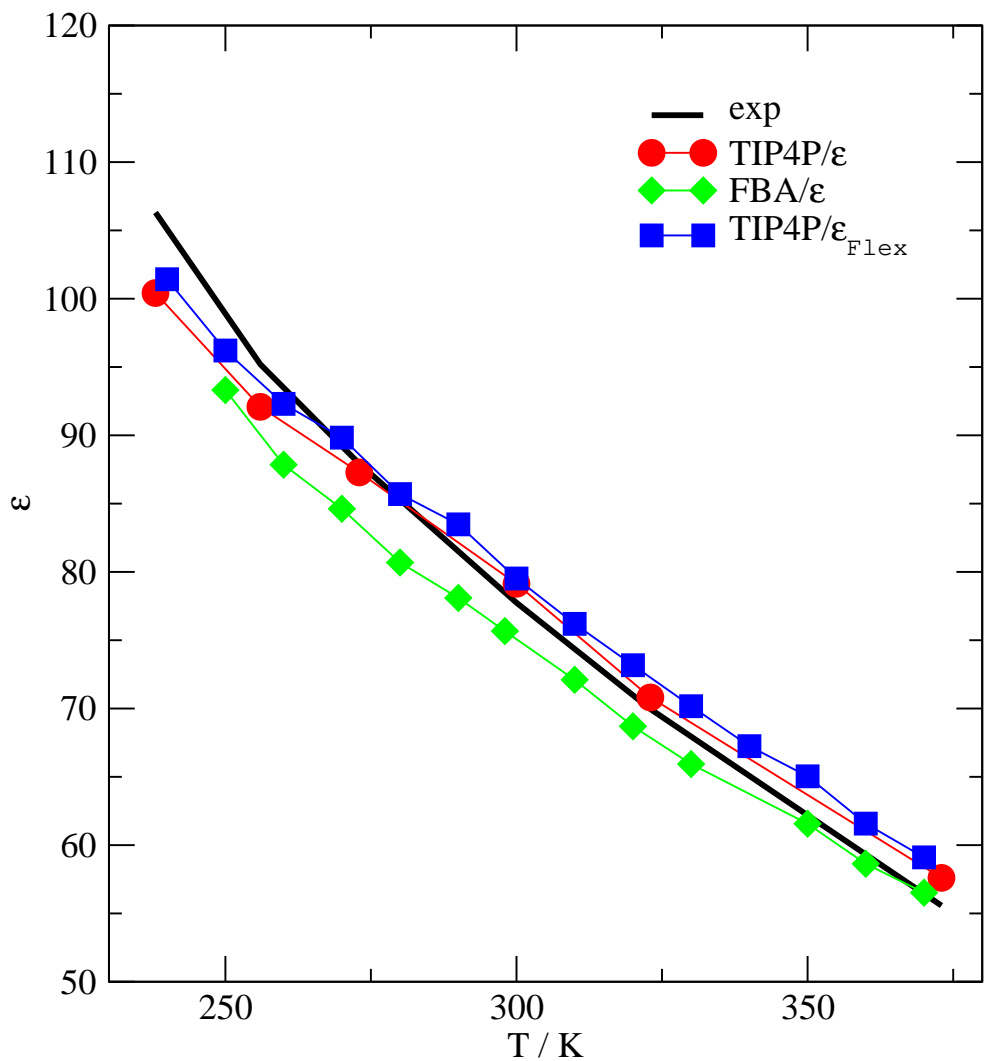}}
\caption{Dielectric constant versus temperature  at $1$ bar of pressure for the TIP4P/$\epsilon\hspace{0.1cm} _{Flex}$(blue squares), FBA/$\epsilon$ (green diamonds) and TIP4P/$\epsilon$ (red circles) models and experimental data~\cite{NIST} (solid line).}
\label{diel}
\end{figure}
\newpage
The proper evaluation of the dielectric constant needs long
simulations to have the average dipole moment of the system
around zero. The dielectric constant
results are shown in figure \ref{diel} at different temperatures and 1 bar of pressure, as the TIP4P/$\epsilon$ the new model TIP4P/$\epsilon\hspace{0.1cm} _{Flex}$ reproduce the experimental values with less error than others models\cite{Piotr}.
The water static dielectric constant, $\epsilon$, is a collective property of an ensemble of water dipoles, which can be calculated from the equilibrium total dipole moment fluctuations, $(<\textbf{M}^2>-<\textbf{M}>^2)$.
The calculations of the dielectric constant was obtained by the equation \ref{Ec4}~\cite{Neumann} of the total dipole moment {\bf M},

\begin{equation}
\label{Ec4}
\epsilon=1+\frac{4\pi}{3k_BTV} (<\textbf{M}^2>-<\textbf{M}>^2)
\end{equation}
\noindent where  $k_B$ is the Boltzmann constant, $T$ is the absolute temperature and $V$ is the volume.


\newpage
\begin{figure}
\centerline{\psfig{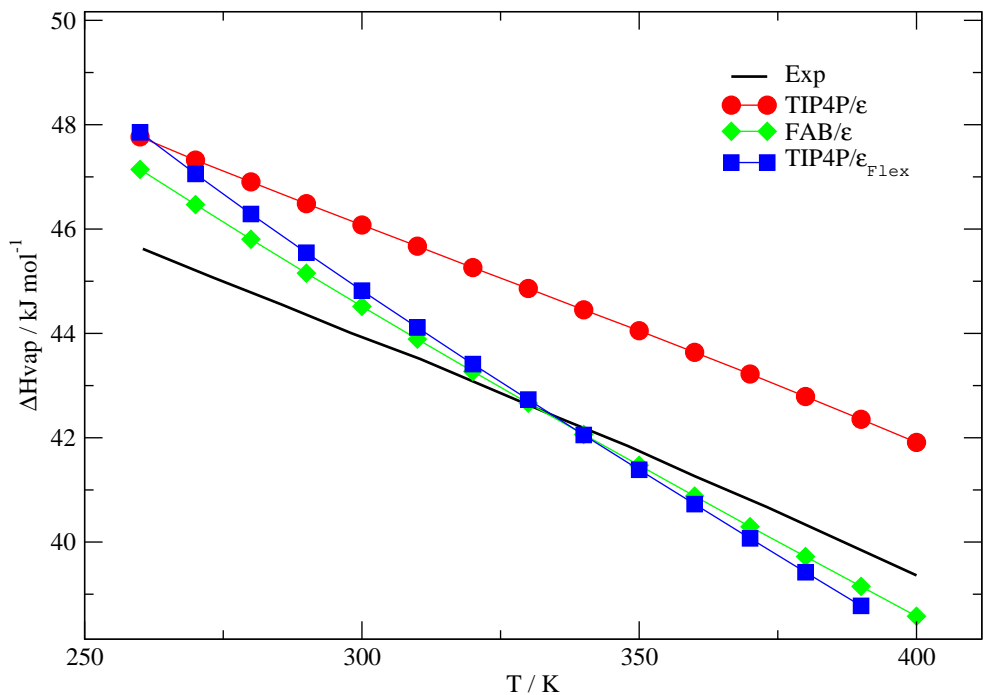}}
\caption{Heat of vaporization $\Delta H_{vap}$
as a function of temperature at pressure constant of 1bar for the TIP4P/$\epsilon\hspace{0.1cm} _{Flex}$(blue squares), FBA/$\epsilon$ (green diamonds) and TIP4P/$\epsilon$ (red circles) models and experimental data~\cite{NIST} (solid line).}
\label{deltaH}
\end{figure}
\newpage
A wide exploration of the properties behavior which is not linked with the parametrization procedure is necessary when a new model is presented. If the model would be robust, even response functions of property of water would be reproducible, functions exhibit a very peculiar behavior in water, and is important to include most of them.\\
Figure~\ref{deltaH}  compares  the heat
of vaporization, $\Delta H_{vap}$, as a function of the temperature at $1\;bar$ for the TIP4P/$\epsilon\hspace{0.1cm} _{Flex}$, FBA/$\epsilon$ and TIP4P/$\epsilon$ water models. It shows that
 flexible FBA/$\epsilon$ and TIP4P/$\epsilon\hspace{0.1cm} _{Flex}$ agree with the data in a range of 298 K to 365 K  with an error less than 2\%, which indicates that this type of model does not need any correction. So the comparison with the non-polarizable models will be without correction, table \ref{table3} and \ref{table4} in this work. The flexibility keeps the correct reproduction for the vapor heat specifically but might affect coexistence properties.

\newpage
\begin{figure}
\centerline{\psfig{figure=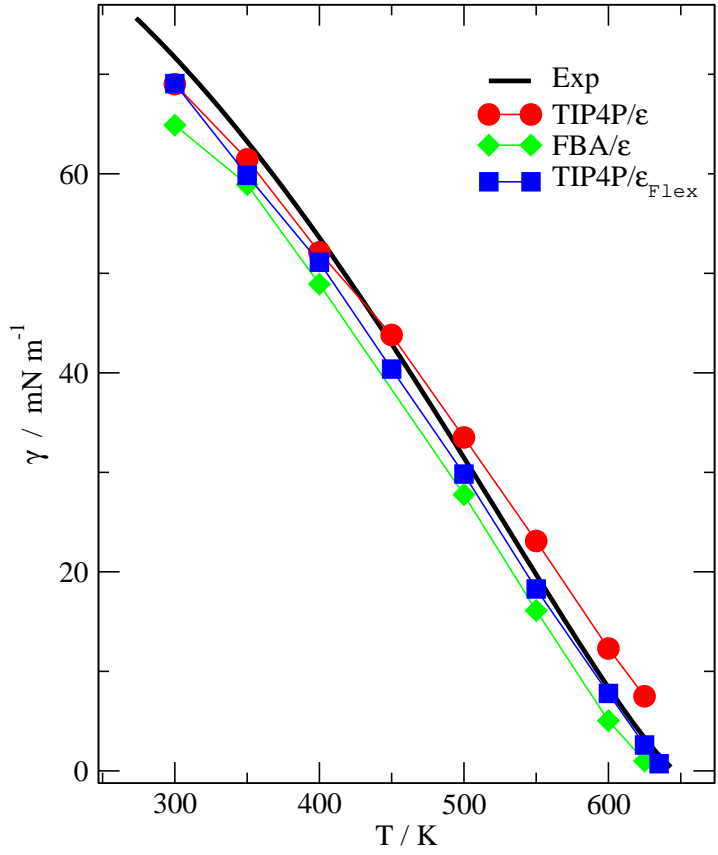,width=14.0cm,angle=-90}}
\caption{Surface tension as a function of temperature  for the TIP4P/$\epsilon\hspace{0.1cm} _{Flex}$(blue squares), FBA/$\epsilon$ (green diamonds) and TIP4P/$\epsilon$ (red circles) models and experimental data~\cite{NIST} (solid line). }
\label{st}
\end{figure}
\newpage
The corresponding surface  tension $\gamma$ on  the planar interface was calculated 
from the mechanical definition of $\gamma$ ~\cite{Lemus}.
\begin{equation}
\gamma=0.5L_z[<P_{zz}>-0.5(<P_{xx}>+<P_{yy}>)]
\end{equation}
where $L_z$ is the length of the simulation cell in the longest direction and $P_{\alpha\alpha}$ are the diagonal components of the pressure 
tensor. The factor 0.5 outside the squared brackets take into account the two symmetrical interfaces in the system.
The surface tension \cite{Lemus} results are shown in Figure \ref{st}. The results for the TIP4P/$\epsilon\hspace{0.1cm} _{Flex}$ are in good agreement
with experimental data \cite{NIST} at all temperatures and improved in high temperatures the TIP4P's results table \ref{table3},\ref{table4}.

\newpage
\begin{figure}
\centerline{\psfig{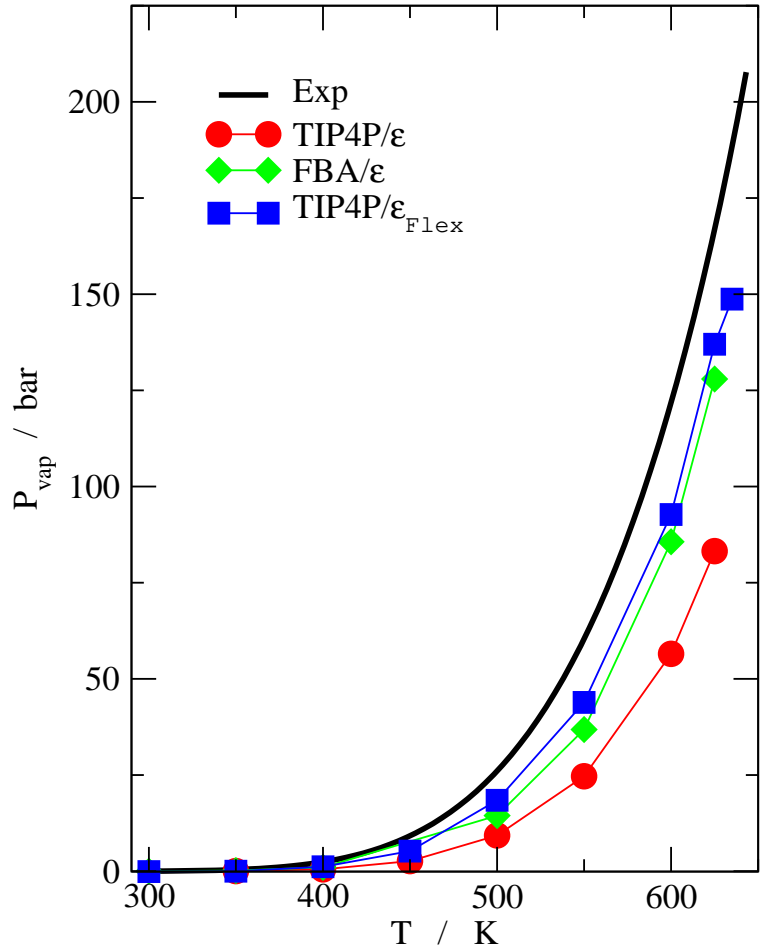}}
\caption{Vapor pressure as a function of temperature for the TIP4P/$\epsilon\hspace{0.1cm} _{Flex}$(blue squares), FBA/$\epsilon$ (green diamonds) and TIP4P/$\epsilon$ (red circles) models and experimental data~\cite{NIST} (solid line). }
\label{pv}
\end{figure}
	\newpage
	The P$_{vap}$ increases logarithmically as the temperature increases and models of 3 and 4 sites rigid non-polarizables can not reproduce this behavior, having great differences with respect to the experimental value, see~ table \ref{table3}, \ref{table4}. The FAB$\epsilon$ model is a little bit closer to the reproduction of the experimental value \cite{fabe}, however the new TIP4P/$\epsilon\hspace{0.1cm} _{Flex}$ model achieves a better reproduction; getting to be counted in the table \ref{table3}. Although the reproduction of this value still fails in all models, the new model is the one that would most be reproduced without compromising the reproduction of the other experimental data.
	
\newpage
\begin{figure}
\centerline{\psfig{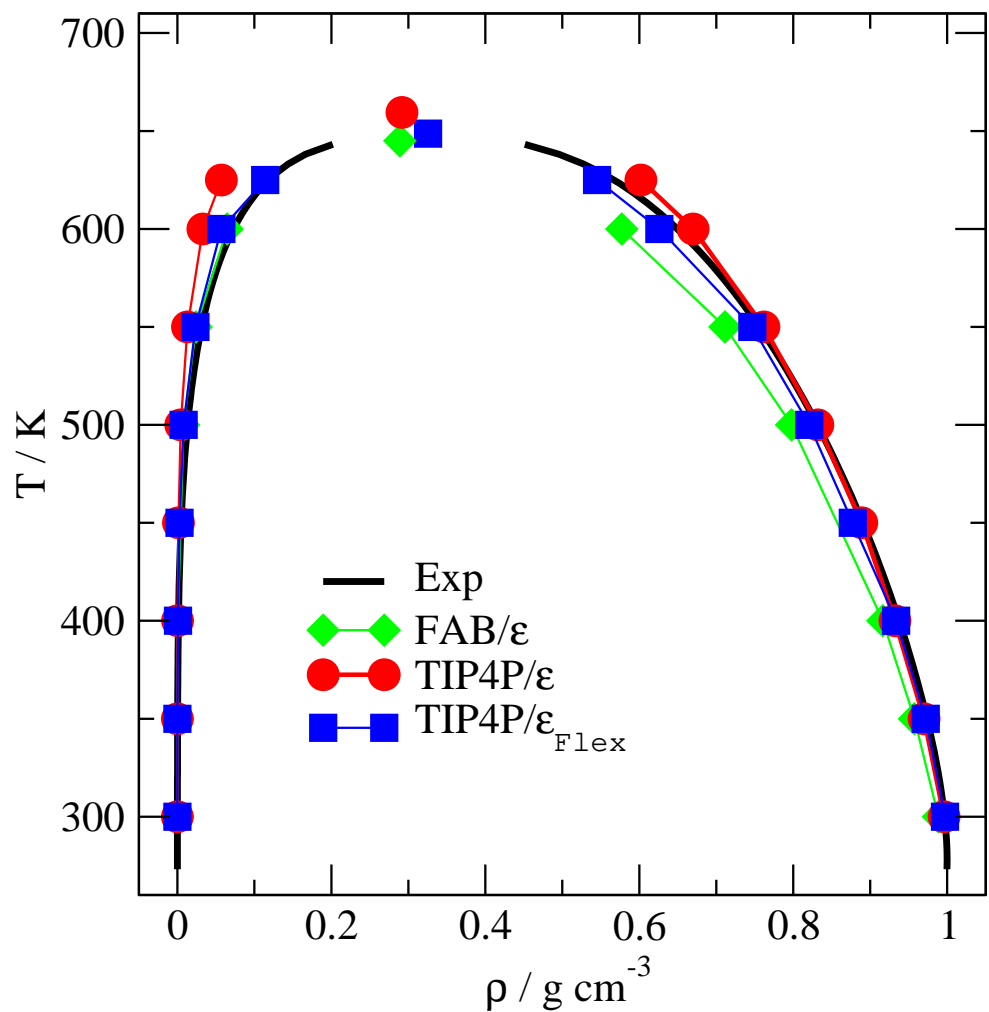}}
\caption{Temperature versus density phase diagram for the  the TIP4P/$\epsilon\hspace{0.1cm} _{Flex}$(blue squares), FBA/$\epsilon$ (green diamonds) and TIP4P/$\epsilon$ (red circles) models and experimental data~\cite{NIST} (solid line).}
\label{lv}
\end{figure}
\newpage
The liquid and vapor densities in simulations of the interface are obtained by fitting the average density profile to a hyperbolic tangent function \cite{Lemus}. The
results shown in figure \ref{lv} and given in table \ref{table3}. The liquid densities predicted
by the TIP4P/$\epsilon\hspace{0.1cm} _{Flex}$ model are in excellent agreement with experimental data \cite{NIST}
at all temperatures.\\

 The model TIP4P/$\epsilon\hspace{0.1cm} _{Flex}$ and FBA/$\epsilon$
give vapor densities close to experiments for all temperatures and their predicted critical temperatures
are 0.24\% closer and 3.06\% smaller than experimental data, respectively. The critical parameters for the TIP4P/$\epsilon\hspace{0.1cm} _{Flex}$ model are $\rho_{C}$ = 0.32562 g $cm^{-3}$ and T$_{C}$=648.66K and they were obtained by using the rectilinear diameters law with critical exponents $\beta$= 0.325.

\newpage
\begin{figure}
\centerline{\psfig{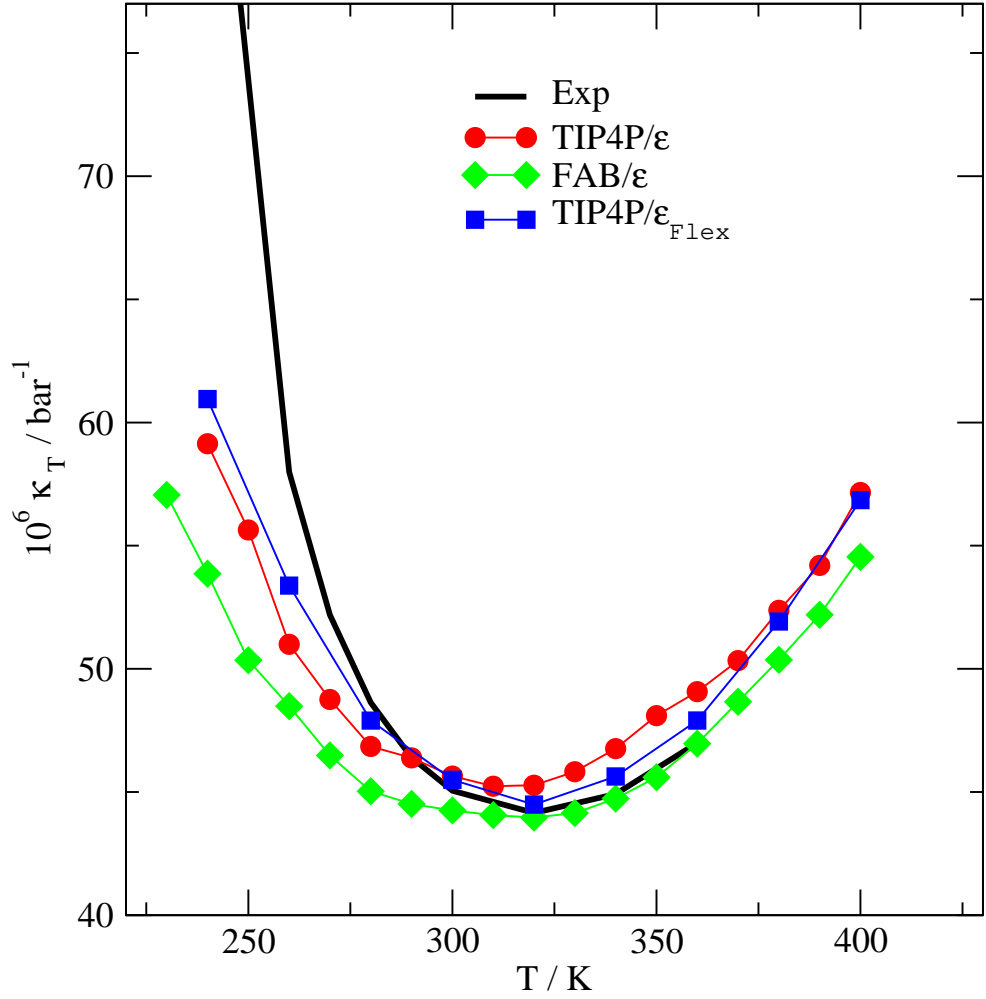}}
\caption{Temperature versus isothermal compressibility 
($\kappa$) for the  the TIP4P/$\epsilon\hspace{0.1cm} _{Flex}$(blue squares), FBA/$\epsilon$ (green diamonds) and TIP4P/$\epsilon$ (red circles) models and experimental data~\cite{NIST} (solid line). }
\label{kapa}
\end{figure}
\newpage
The values of $\kappa_T$ were obtained from simulations on the isothermal-isobaric ensemble, using

\begin{equation}
\kappa_T = -\frac{1}{V}\left(\frac{\partial V}{\partial P}\right)_T = \frac{1}{\rho} \left ( \frac {\partial \rho} {\partial P} \right )_T 
\end{equation}

Simulations of systems containing 500 molecules are performed at 9 different temperatures: T = 240~K,
260~K, 280~K, 300~K 320~K, 340~K, 360~K, 380~K and 400~K. At each temperature, the pressure is varied from $P = 1$~bar to $P = 1200$~bars. 
The density as a function of pressure is fitted with a cubic polynomial at 
each temperature and the value of $\kappa_T$ at $P = 1$ bar is determined from the respective polynomial. The results are given in table \ref{table3},\ref{table3} and  in  figure \ref{kapa}. The TIP4P/$\epsilon\hspace{0.1cm} _{Flex}$ model gives a minimum value of $\kappa_T$ at the same postion of experimental data. In general TIP4P/$\epsilon\hspace{0.1cm} _{Flex}$ model gives results closer to experiment than the TIP4P/$\epsilon$ and FBA/$\epsilon$. 
\newpage
\begin{figure}
\centerline{\psfig{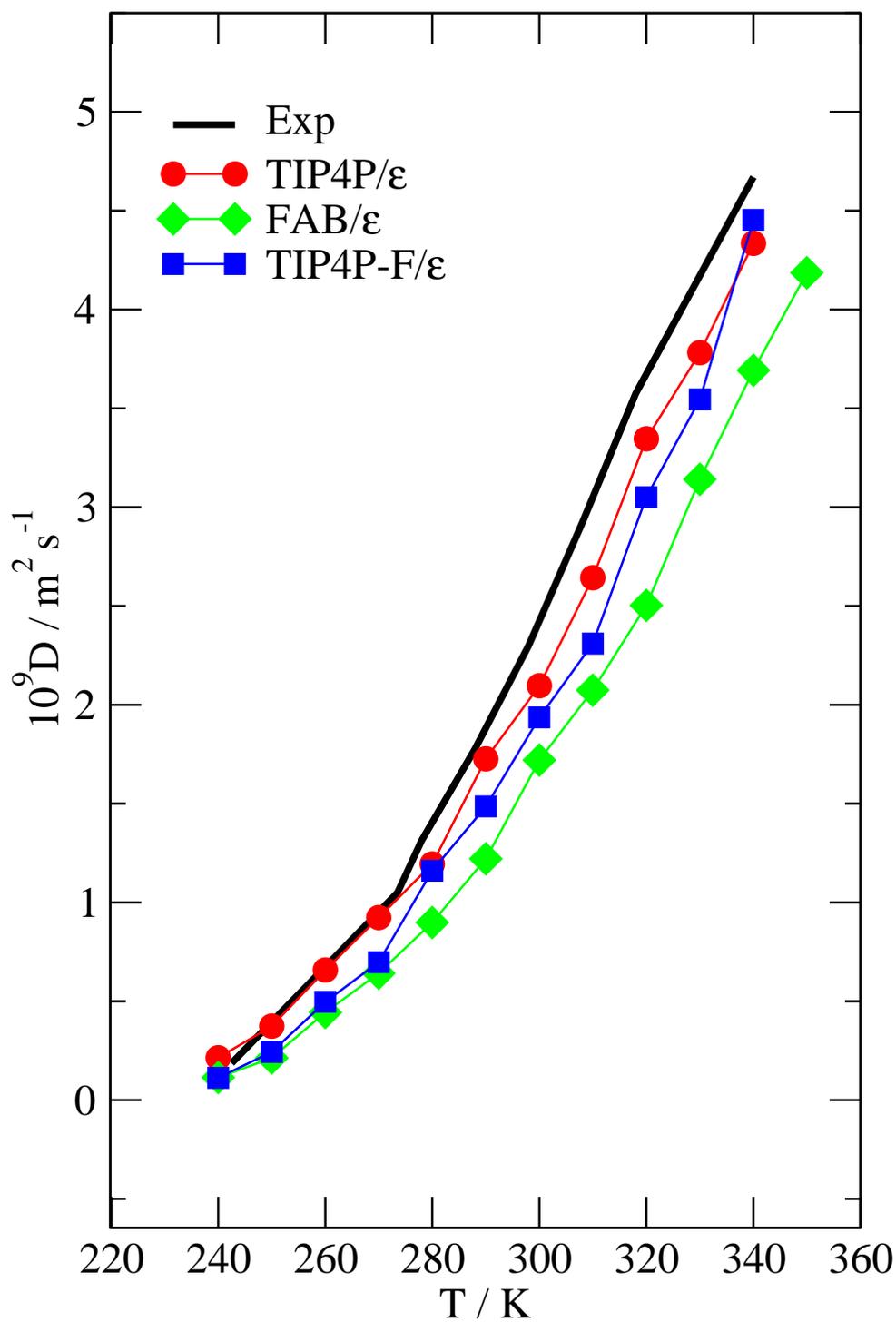}}
\caption{Temperature versus self-diffusion coefficient 
(D) for the  the TIP4P/$\epsilon\hspace{0.1cm} _{Flex}$(blue squares), FBA/$\epsilon$ (green diamonds) and TIP4P/$\epsilon$ (red circles) models and experimental data~\cite{NIST} (solid line). }
\label{D}
\end{figure}
\newpage
It  is obtained from the Einstein equation

\begin{equation}
D= \lim_{t \rightarrow \infty} \frac{1}{6t}\left<|{\bf R}_i(t)-{\bf R}_i(0)|^2\right>,
\end{equation}

\noindent where ${\bf R}_i(t)$ is the center of mass position of molecule $i$ at time $t$ and $< ... >$ denotes time average.

The diffusion coefficient of TIP4P/$\epsilon\hspace{0.1cm} _{Flex}$  is shown as a function of the temperature in figure \ref{D} and given in table \ref{table3}, as well as the non-polarizable rigid water models.

\newpage
\begin{figure}
\centerline{\psfig{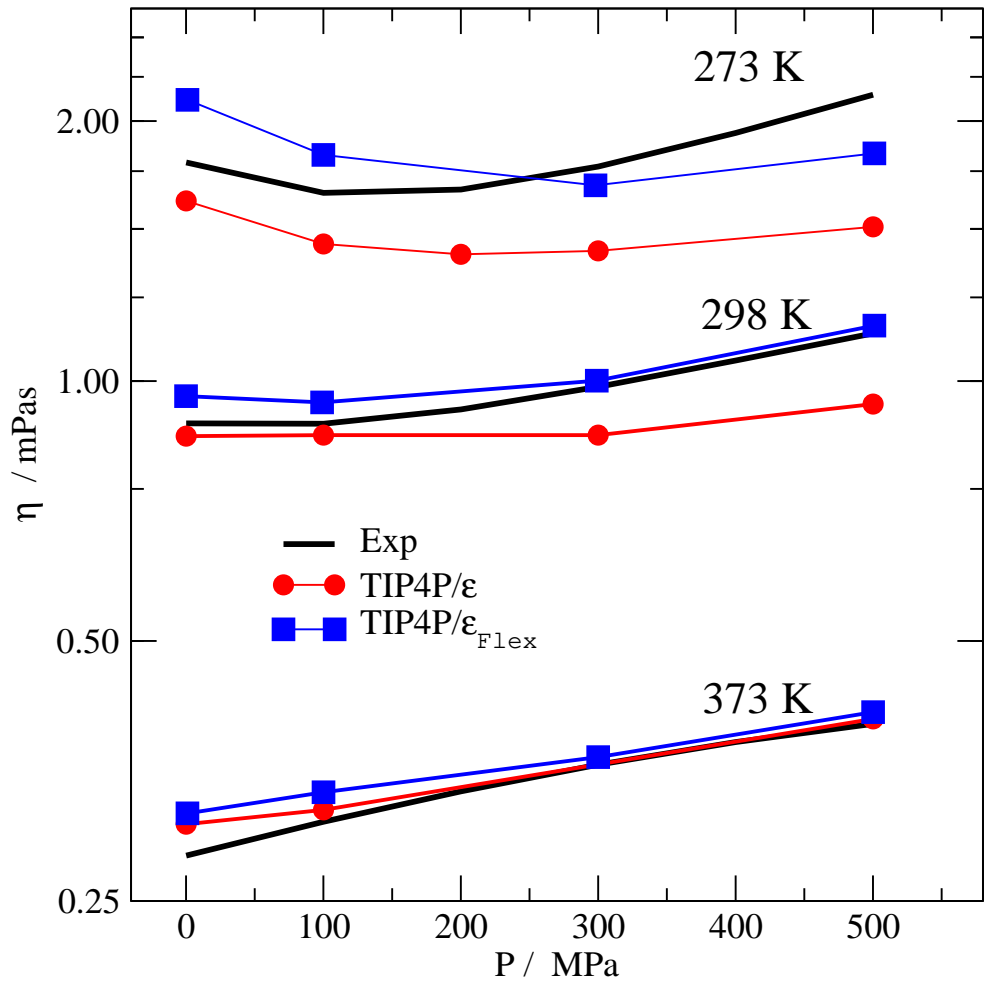}}
\caption{Temperature versus shear viscosity 
($\eta$) for the  the TIP4P/$\epsilon\hspace{0.1cm} _{Flex}$(blue squares) and TIP4P/$\epsilon$ (red circles) models and experimental data~\cite{NIST} (solid line). }
\label{visc}
\end{figure}
\newpage
The shear viscosity was obtained in this work for all models in tables \ref{table3} and \ref{table4}, NPT simulations are performed for at least 20
million steps with a time step of 1 fs\cite{visc}.
The pressure components are saved on a disk every 1 fs, and all the configurations are used as a time origin. The upper limit in the integrations is 8 ps in all cases. The results are shown in figure \ref{visc} for The TIP4P/$\epsilon\hspace{0.1cm} _{Flex}$ and TIP4P/$\epsilon$ at different temperatures and pressures and in tables \ref{table3} and \ref{table4} for the rest of the models considered. The TIP4P/$\epsilon\hspace{0.1cm} _{Flex}$ and TIP4P/$\epsilon$ give the same values, within the simulation
error at  373K. But at 273 K, the simulation values  are systematically lower for all pressures for TIP4P/$\epsilon$; at 298K gives the same values at lower pressures but the difference is larger at high pressures.
Both models predict the minimum value of shear viscosity at 273 K and 298K.

\newpage
\begin{figure}
\centerline{\psfig{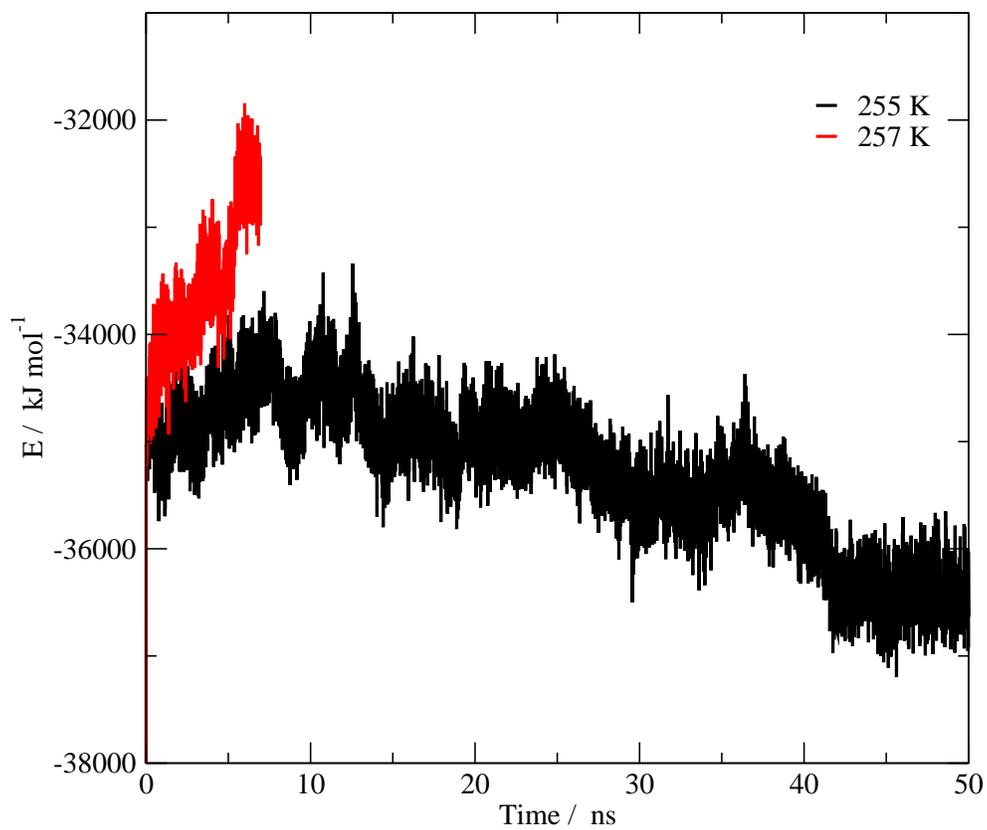}}
\caption{Total energy of a water system vs time, betwen liquid water in contact with ice Ih for the TIP4P/$\epsilon\hspace{0.1cm} _{Flex}$ model. The results are  NpT simulation runs at 1 bar and T = 250 K and 252 K. }
\label{tfus}
\end{figure}
\newpage
The TIP4P/$\epsilon\hspace{0.1cm} _{Flex}$ force field reproduces the melting temperature, T${_m}$ with a difference of 6.59\% with respect to the experimental data. Figure\ref{tfus}, the red line describes the behavior of the total energy of the system when it turns into a liquid phase and the black line has an energetic decrease that indicates that there is an entire Ice Ih-transforming system. The result of T${_m}$
is estimated to be 255.5 K figure\ref{tfus}. This result improved the value of TIP4P/$\epsilon$ in 15.5 degrees of temperature.\\
To calculate this property using direct
coexistence simulations.\cite{melting} Anisotropic NPT simulations, starting from the same initial configuration, at 1 bar and temperatures ranging from 250 to 260 K with one degree of temperature in each simulation. Simulation was carried out on systems containing 870 water molecules in an elongated simulation cell in the z direction, half in the liquid and half in the ice Ih. The procedure is observing the melting or freezing by inspecting the total energy, if the system is above the T${_m}$ the
ice region will melt, but if the water is at a temperature
below the melting point the liquid water will be ice Ih \cite{melting}.

In table \ref{table3} and \ref{table4} compared the most successful model of 3 and 4 sites with respect to the experimental data that can reproduce follow the procedure of Vega et al \cite{Vega11} for a certain property the prediction of a given water model adopts the value X, and the experimental value is X$_{exp}$. The score is calculated by the equation (4.8)
\begin{equation}
M= min\left\{anint\left[10 -  abs\left(\frac{X-X_{exp}}{X_{exp}   \textit{tol}}\right)\right], 0 \right\},
\end{equation}
where the tolerance \textit{tol} is given as a percentage and anint is the nearest integer function.

\newpage
\begin{table}
\caption{Experimental and simulation data of TIP4P/05,	TIP4P/$\epsilon$, FAB/$\epsilon$, TIP4P/$\epsilon\hspace{0.1cm} _{Flex}$, water models. Thermodynamic conditions as reported in each entry.}
\label{table3}
\scalebox{0.7}[0.65]{
\begin{tabular}{|lcccccccccc|}
\hline\hline
	&		&		&		&		&		&		&		&		&		&		\\
Property	&	Exp.	&	TIP4P/05	&	TIP4P/$\epsilon$	&	FAB/$\epsilon$	&	TIP4P/$\epsilon\hspace{0.1cm} _{Flex}$	&	Tol. (\%)	&	TIP4P/05	&	TIP4P/$\epsilon$	&	FAB/$\epsilon$	&	TIP4P/$\epsilon\hspace{0.1cm} _{Flex}$	\\
	&	data	&		&		&		&		&		&	score	&	score	&	score	&	score	\\
\hline																					
\multicolumn{11}{|l|}{Enthalpy of phase change / kcal mol$^{-1}$}\\																					
\hline																					
$\Delta$H$_{melt}$	&	1.44	&	1.16	&	1.22	&	0.882	&	1.799	&	5	&	6.1	&	6.9	&	2.3	&	5.0	\\
$\Delta$H$_{vap}$	&	10.52	&	11.99	&	11.08	&	10.74	&	10.84	&	5	&	7.2	&	8.9	&	9.6	&	9.4	\\
\hline																					
\multicolumn{11}{|l|}{Critical point properties}\\																					
\hline																					
T$_C$/K	&	647.1	&	640	&	659.5	&	627.28	&	648.66	&	1	&	8.9	&	8.1	&	6.9	&	9.8	\\
$\rho_C$/g cm${^{-3}}$	&	0.322	&	0.31	&	0.29512	&	0.2854	&	0.32567	&	1	&	6.3	&	1.7	&	0.0	&	8.9	\\
\hline																					
\multicolumn{11}{|l|}{Surface tension/mN m$^{-1}$}\\																					
\hline																					
$\gamma_{300K}$	&	71.73	&	69.09	&	69.00	&	64.88	&	69.05	&	2.5	&	8.5	&	8.5	&	6.2	&	8.5	\\
$\gamma_{400K}$	&	53.6	&	50.60	&	52.11	&	48.91	&	51.10	&	2.5	&	7.8	&	8.9	&	6.5	&	8.1	\\
$\gamma_{600K}$	&	8.4	&	7.63	&	7.86	&	5.05	&	7.78	&	2.5	&	6.3	&	7.4	&	0.0	&	7.1	\\
\hline																					
	\multicolumn{11}{|l|}{Melting properties}\\																				
\hline																					
T$_m$/K	&	273.15	&	252	&	240	&	251	&	256	&	2.5	&	6.9	&	5.1	&	6.8	&	7.5	\\
$\rho_{liq}$/g cm$^{-3}$	&	0.999	&	0.993	&	0.997	&	0.9946	&	0.989784	&	0.5	&	8.8	&	9.6	&	9.1	&	8.2	\\
$\rho_{sol}$/g cm$^{-3}$	&	0.917	&	0.921	&	0.929	&	0.93	&	0.942507	&	0.5	&	9.1	&	7.4	&	7.2	&	4.4	\\
\hline																					
\multicolumn{11}{|l|}{Orthobaric densities and temperature of maximun density \textbf{TMD}}\\																					
\hline																					
\textbf{TMD}/K	&	277	&	278	&	277	&	275.4	&	277	&	1	&	9.6	&	10.0	&	9.4	&	10.0	\\
$\rho_{260K}$/g cm$^{-3}$	&	0.9969	&	0.9970	&	0.9980	&	0.9964	&	0.9973	&	0.5	&	10.0	&	9.8	&	9.9	&	9.9	\\
$\rho_{298K}$/g cm$^{-3}$	&	0.9970	&	0.9930	&	0.9963	&	0.9949	&	0.9970	&	0.5	&	9.2	&	9.9	&	9.6	&	10.0	\\
$\rho_{400K}$/g cm$^{-3}$	&	0.9375	&	0.9300	&	0.9368	&	0.9405	&	0.9361	&	0.5	&	8.4	&	9.9	&	9.4	&	9.7	\\
$\rho_{450K}$/g cm$^{-3}$	&	0.8903	&	0.8790	&	0.8842	&	0.8982	&	0.8801	&	0.5	&	7.5	&	8.6	&	8.2	&	7.7	\\
\hline																					
	\multicolumn{11}{|l|}{Isothermal compressibility / 10$^-6$ bar$^{-1}$)}\\																				
\hline																					
$\kappa_T$ [1 bar; 298 K]	&	45.3	&	46.85	&	45.77	&	44.29	&	45.7	&	1	&	6.6	&	9.0	&	7.8	&	9.1	\\
$\kappa_T$ [1 bar;318 K]	&	44.25	&	46.46	&	45.25	&	43.97	&	44.58	&	1	&	5.0	&	7.7	&	9.4	&	9.3	\\
$\kappa_T$ [1 bar;360 K]	&	47	&	50.17	&	49.01	&	46.89	&	47.901	&	1	&	3.3	&	5.7	&	9.8	&	8.1	\\
\hline																					
\multicolumn{11}{|l|}{Thermal expansion coefficient   / 10$^5$ K$^{-1}$)}\\																					
\hline																					
$\alpha_P$ [1 bar; 298 K]	&	22.66	&	22.6	&	23.79	&	29.29	&	24.55	&	5	&	9.9	&	9.0	&	4.1	&	8.3	\\
$\alpha_P$ [1 bar;350 K]	&	68.2	&	69.54	&	70.87	&	65.83	&	65.06	&	5	&	9.6	&	9.2	&	9.3	&	9.1	\\
\hline																					
\multicolumn{11}{|l|}{Heat capacity at constant pressure/    cal mol $^{-1}$K$^{-1}$}\\																					
\hline																					
$C_{p}$ [liq 298 K; 1 bar]	&	18	&	21.10	&	19.80	&	25.20	&	28.00	&	5	&	6.6	&	8.0	&	2.0	&	0.0	\\
$C_{p}$ [ice 250 K; 1 bar]	&	8.3	&	14.00	&	13.90	&	7.65	&	21.00	&	5	&	0.0	&	0.0	&	8.4	&	0.0	\\
\hline																					
\multicolumn{11}{|l|}{Liquid-vapor phase equilibrium (density)  /  g $ cm^{-3}$ }\\																					
\hline																					
$\rho_{vap}$[500 K]	&	0.0128	&	0.006	&	0.0044	&	0.0078	&	0.0082	&	5	&	0.0	&	0.0	&	2.2	&	2.8	\\
$\rho_{vap}$[625 K]	&	0.1139	&	0.13	&	0.057	&	0.161	&	0.1141	&	5	&	7.2	&	0.0	&	1.7	&	10.0	\\
$\rho_{liq}$[400 K] 	&	0.9433	&	0.932	&	0.932	&	0.9163	&	0.9337	&	5	&	9.8	&	9.8	&	9.4	&	9.8	\\
$\rho_{liq}$[500 K]	&	0.6694	&	0.821	&	0.832	&	0.7977	&	0.8209	&	5	&	5.5	&	5.1	&	6.2	&	5.5	\\
$\rho_{liq}$[625 K]	&	0.5764	&	0.54	&	0.602	&	0.489	&	0.5452	&	5	&	8.7	&	9.1	&	7.0	&	8.9	\\
\hline																					
\multicolumn{11}{|l|}{Gas properties    /    bar }\\																					
\hline																					
$p_{v}$[450 K] 	&	9.32	&	4.46	&	2.64	&	4.29	&	5.23	&	5	&	0.0	&	0.0	&	0.0	&	1.2	\\
$p_{v}$[550 K] 	&	60.48	&	16.1	&	24.69	&	36.83	&	43.9	&	5	&	0.0	&	0.0	&	2.2	&	4.5	\\
\hline																					
\multicolumn{11}{|l|}{Static dielectric constant}\\																					
\hline																					
$\varepsilon$[liq; 298 K]	&	78.5	&	58	&	79.36	&	75.65	&	79.06	&	1	&	0.0	&	8.9	&	6.4	&	9.3	\\
$\varepsilon$[liq; 350 K]	&	62.12	&	43.1	&	63.51	&	61.56	&	64.93	&	1	&	0.0	&	7.8	&	9.1	&	5.5	\\
$\varepsilon$[10kbar,300K]	&	103.63	&	81.9	&	103.2	&	104.9	&	102.296	&	1	&	0.0	&	9.6	&	8.8	&	8.7	\\
\hline																					
\multicolumn{11}{|l|}{T$_m$-TMD-T$_c$. ratios}\\																					
\hline																					
T$_m$[I$_h$]/T$_c$	&	0.422	&	0.394	&	0.36	&	0.40	&	0.39	&	5	&	8.7	&	7.2	&	9.0	&	8.7	\\
TMD/T$_c$	&	0.428	&	0.434	&	0.42	&	0.43	&	0.42	&	5	&	9.7	&	9.6	&	9.5	&	10.0	\\
TMD-T$_m$(K)	&	4	&	26	&	37	&	24.4	&	21	&	5	&	0.0	&	0.0	&	0.0	&	0.0	\\
\hline																					
\multicolumn{11}{|l|}{Densities of ice Ih/g cm$^{-3}$}\\																					
\hline																					
$\rho$[I$_h$ 250 K; 1 bar]	&	0.92	&	0.9207	&	0.922	&	0.93	&	0.939	&	0.5	&	9.8	&	9.6	&	7.8	&	5.9	\\
$\rho$[I$_h$ 220 K; 1 bar]	&	0.923	&	0.9248	&	0.926	&	0.935	&	0.94	&	0.5	&	9.6	&	9.3	&	7.4	&	6.3	\\
$\rho$[I$_h$ 150 K; 1 bar]	&	0.969	&	0.936	&	0.937	&	0.974	&	0.979	&	0.5	&	3.2	&	3.4	&	9.0	&	7.9	\\
\hline																					
\multicolumn{11}{|l|}{EOS high pressure}\\																					
\hline																					
$\rho$[373 K; 10 kbar]	&	1.201	&	1.204	&	1.202	&	1.215	&	1.1977	&	0.5	&	9.5	&	9.9	&	7.7	&	9.5	\\
$\rho$[373 K; 20 kbar]	&	1.322	&	1.321	&	1.318	&	1.339	&	1.3131	&	0.5	&	9.8	&	9.3	&	7.4	&	8.7	\\
\hline																					
\multicolumn{11}{|l|}{Self-diffusion coefficient/cm$^2$s$^{-1}$ }\\																					
\hline																					
ln D$_{278K}$	&	-11.24	&	-11.27	&	-11.27	&	-11.58	&	-11.46	&	1	&	9.7	&	9.7	&	7.0	&	8.0	\\
ln D$_{298K}$	&	-10.68	&	-10.79	&	-10.79	&	-11.01	&	-10.908	&	1	&	9.0	&	9.0	&	6.9	&	7.9	\\
E$_a$ k$_J$ mol$^{ -1}$	&	18.4	&	16.2	&	20.51	&	15.98	&	19	&	2	&	4.0	&	4.3	&	3.4	&	8.4	\\
\hline																					
\multicolumn{11}{|l|}{Shear viscosity / mPa s}\\																					
\hline																					
$\eta$[1 bar; 298 K]	&	0.896	&	0.855	&	0.855	&	1.265	&	0.958	&	5	&	9.1	&	9.1	&	1.8	&	8.6	\\
$\eta$[1 bar; 373 K]	&	0.284	&	0.289	&	0.289	&	0.364	&	0.314	&	5	&	9.6	&	9.6	&	4.4	&	7.9	\\
\hline																					
\hline																					
\multicolumn{7}{|l}{Overall score (out of 10)}													&	6.15	&	6.81	&	6.62	&	7.24	\\

\hline
\end{tabular}}
\end{table}

\newpage

\begin{table}
\caption{Experimental and simulation data of TIP4P-FB,	TIP4P-ST and OPC water models. Thermodynamic conditions as reported in each entry.•}
\label{table4}
\scalebox{0.8}[0.65]{
\begin{tabular}{|lcccccccc|}
\hline\hline

	&		&		&		&		&		&		&		&		\\
Property	&	Exp.	&	TIP4P-FB	&	TIP4P-ST	&	OPC	&	Tol. (\%)	&	TIP4P-FB	&	TIP4P-ST	&	OPC	\\
	&	data	&		&		&		&		&	score	&	score	&	score	\\
\hline																	
\multicolumn{9}{|l|}{Enthalpy of phase change / kcal mol$^{-1}$}\\																	
\hline																	
$\Delta$H$_{melt}$	&	1.44	&	1.24	&	1.24	&	0.871	&	5	&	7.2	&	7.2	&	2.1	\\
$\Delta$H$_{vap}$	&	10.52	&	10.9	&	10.9	&	11.06	&	5	&	9.3	&	9.3	&	9.0	\\
\hline																	
\multicolumn{9}{|l|}{Critical point properties}\\																	
\hline																	
T$_C$/K	&	647.1	&	658.5	&	660.5	&	685	&	1	&	8.2	&	7.9	&	4.1	\\
$\rho_C$/g cm${^{-3}}$	&	0.322	&	0.2983	&	0.2970&	0.3019	&	1	&	2.7	&	2.3	&	3.8	\\
\hline																	
\multicolumn{9}{|l|}{Surface tension/mN m$^{-1}$}\\																	
\hline																	
$\gamma_{300K}$	&	71.73	&	67.13	&	68.98	&	74.44	&	2.5	&	7.4	&	8.5	&	8.5	\\
$\gamma_{400K}$	&	53.6	&	53.76	&	52.07	&	58.81	&	2.5	&	9.9	&	8.9	&	6.1	\\
$\gamma_{600K}$	&	8.4	&	12.30	&	11.82	&	18.78	&	2.5	&	0.0	&	0.0	&	0.0	\\
\hline																	
	\multicolumn{9}{|l|}{Melting properties}\\																
\hline																	
T$_m$/K	&	273.15	&	243	&	246	&	242	&	2.5	&	5.6	&	6.0	&	5.4	\\
$\rho_{liq}$/g cm$^{-3}$	&	0.999	&	0.988	&	0.99	&	0.894	&	0.5	&	7.8	&	8.2	&	0.0	\\
$\rho_{sol}$/g cm$^{-3}$	&	0.917	&	0.926	&	0.928	&	0.994	&	0.5	&	8.0	&	7.6	&	0.0	\\
\hline																	
\multicolumn{9}{|l|}{Orthobaric densities and temperature of maximun density \textbf{TMD}}\\																	
\hline																	
\textbf{TMD}/K	&	277	&	281	&	277	&	272	&	1	&	8.6	&	10.0	&	8.2	\\
$\rho_{260K}$/g cm$^{-3}$	&	0.9969	&	0.9950	&	0.9980	&	1.0000	&	0.5	&	9.6	&	9.8	&	9.4	\\
$\rho_{298K}$/g cm$^{-3}$	&	0.9970	&	0.9958	&	0.9970	&	0.9966	&	0.5	&	9.8	&	10.0	&	9.9	\\
$\rho_{400K}$/g cm$^{-3}$	&	0.9375	&	0.9384	&	0.9396	&	0.9400	&	0.5	&	9.8	&	9.6	&	9.5	\\
$\rho_{450K}$/g cm$^{-3}$	&	0.8903	&	0.8914	&	0.8911	&	0.8984	&	0.5	&	9.8	&	9.8	&	8.2	\\
\hline																	
	\multicolumn{9}{|l|}{Isothermal compressibility / 10$^-6$ bar$^{-1}$)}\\																
\hline																	
$\kappa_T$ [1 bar; 298 K]	&	45.3	&	44.8	&	45.2	&	47.4	&	1	&	8.9	&	9.8	&	5.4	\\
$\kappa_T$ [1 bar;318 K]	&	44.25	&	44.9	&	45.2	&	46.6	&	1	&	8.5	&	7.9	&	4.7	\\
$\kappa_T$ [1 bar;360 K]	&	47	&	49.4	&	48.93	&	48.5	&	1	&	4.9	&	5.9	&	6.8	\\
\hline																	
\multicolumn{9}{|l|}{Thermal expansion coefficient   / 10$^5$ K$^{-1}$)}\\																	
\hline																	
$\alpha_P$ [1 bar; 298 K]	&	22.66	&	23	&	24	&	25.62	&	5	&	9.7	&	8.8	&	7.4	\\
$\alpha_P$ [1 bar;350 K]	&	68.2	&	61.9	&	60.3	&	67.49	&	5	&	8.2	&	7.7	&	9.8	\\
\hline																	
\multicolumn{9}{|l|}{Heat capacity at constant pressure/    cal mol $^{-1}$K$^{-1}$}\\																	
\hline																	
$C_{p}$ [liq 298 K; 1 bar]	&	18	&	19.20	&	18.90	&	18.00	&	5	&	8.7	&	9.0	&	10.0	\\
$C_{p}$ [ice 250 K; 1 bar]	&	8.3	&	13.93	&	14.03	&	13.81	&	5	&	0.0	&	0.0	&	0.0	\\
\hline																	
\multicolumn{9}{|l|}{Liquid-vapor phase equilibrium (density)  / g  $ cm^{-3}$ }\\																	
\hline																	
$\rho_{vap}$[500 K]	&	0.0128	&	0.005	&	0.0045	&	0.0017	&	5	&	0.0	&	0.0	&	0.0	\\
$\rho_{vap}$[625 K]	&	0.1139	&	0.06059	&	0.058	&	0.034	&	5	&	0.6	&	0.2	&	0.0	\\
$\rho_{liq}$[400 K] 	&	0.9433	&	0.938	&	0.936	&	0.938	&	5	&	9.9	&	9.8	&	9.9	\\
$\rho_{liq}$[500 K]	&	0.6694	&	0.831	&	0.833	&	0.846	&	5	&	5.2	&	5.1	&	4.7	\\
$\rho_{liq}$[625 K]	&	0.5764	&	0.597	&	0.602	&	0.657	&	5	&	9.3	&	9.1	&	7.2	\\
\hline																	
\multicolumn{9}{|l|}{Gas properties    /    bar }\\																	
\hline																	
$p_{v}$[450 K] 	&	9.32	&	3.102&	2.86374	&	3.1	&	5	&	0.0	&	0.0	&	0.0	\\
$p_{v}$[550 K] 	&	60.48	&	25.042	&	22.30	&	43.3	&	5	&	0.0	&	0.0	&	4.3	\\
\hline																	
\multicolumn{9}{|l|}{Static dielectric constant}\\																	
\hline																	
$\varepsilon$[liq; 298 K]	&	78.5	&	76.86	&	82	&	79.5	&	1	&	7.9	&	5.5	&	8.7	\\
$\varepsilon$[liq; 350 K]	&	62.12	&	62.6	&	64.3	&	63.178	&	1	&	9.2	&	6.5	&	8.3	\\
$\varepsilon$[10kbar,300K]	&	103.63	&	100.874	&	100.596	&	98.413	&	1	&	7.3	&	7.1	&	5.0	\\
\hline																	
\multicolumn{9}{|l|}{T$_m$-TMD-T$_c$. ratios}\\																	
\hline																	
T$_m$[I$_h$]/T$_c$	&	0.422	&	0.36	&	0.37	&	0.35	&	5	&	7.5	&	7.7	&	6.7	\\
TMD/T$_c$	&	0.428	&	0.42	&	0.41	&	0.39	&	5	&	9.9	&	9.6	&	8.6	\\
TMD-T$_m$(K)	&	4	&	38	&	31	&	30	&	5	&	0.0	&	0.0	&	0.0	\\
\hline																	
\multicolumn{9}{|l|}{Densities of ice Ih/g cm$^{-3}$}\\																	
\hline																	
$\rho$[I$_h$ 250 K; 1 bar]	&	0.92	&	0.925	&	0.927	&	0.893	&	0.5	&	8.9	&	8.5	&	4.1	\\
$\rho$[I$_h$ 220 K; 1 bar]	&	0.923	&	0.929	&	0.932	&	0.897	&	0.5	&	8.7	&	8.0	&	4.4	\\
$\rho$[I$_h$ 150 K; 1 bar]	&	0.969	&	0.939	&	0.942	&	0.906	&	0.5	&	3.8	&	4.4	&	0.0	\\
\hline																	
\multicolumn{9}{|l|}{EOS high pressure}\\																	
\hline																	
$\rho$[373 K; 10 kbar]	&	1.201	&	1.205	&	1.206	&	1.18777	&	0.5	&	9.3	&	9.2	&	7.8	\\
$\rho$[373 K; 20 kbar]	&	1.322	&	1.322	&	1.324	&	1.29752	&	0.5	&	10.0	&	9.7	&	6.3	\\
\hline																	
\multicolumn{9}{|l|}{Self-diffusion coefficient/cm$^2$s$^{-1}$ }\\																	
\hline																	
ln D$_{278K}$	&	-11.24	&	-11.16	&	-11.25	&	-11.31	&	1	&	9.3	&	9.9	&	9.4	\\
ln D$_{298K}$	&	-10.68	&	-10.65	&	-10.66	&	-10.8	&	1	&	9.7	&	9.8	&	8.9	\\
E$_a$ k$_J$ mol$^{ -1}$	&	18.4	&	17.56	&	20.318	&	17.56	&	2	&	7.7	&	4.8	&	7.7	\\
\hline																	
\multicolumn{9}{|l|}{Shear viscosity / mPa s}\\																	
\hline																	
$\eta$[1 bar; 298 K]	&	0.896	&	0.95	&	0.81	&	0.815	&	5	&	8.8	&	8.1	&	8.2	\\
$\eta$[1 bar; 373 K]	&	0.284	&	0.287	&	0.283	&	0.277	&	5	&	9.8	&	9.9	&	9.5	\\
\hline																	
\hline																	
\multicolumn{6}{|l}{Overall score (out of 10)}											&	6.6	&	6.4	&	5.4	\\

\hline
\end{tabular}}
\end{table}

\newpage

\begin{figure}
\centerline{\psfig{figure=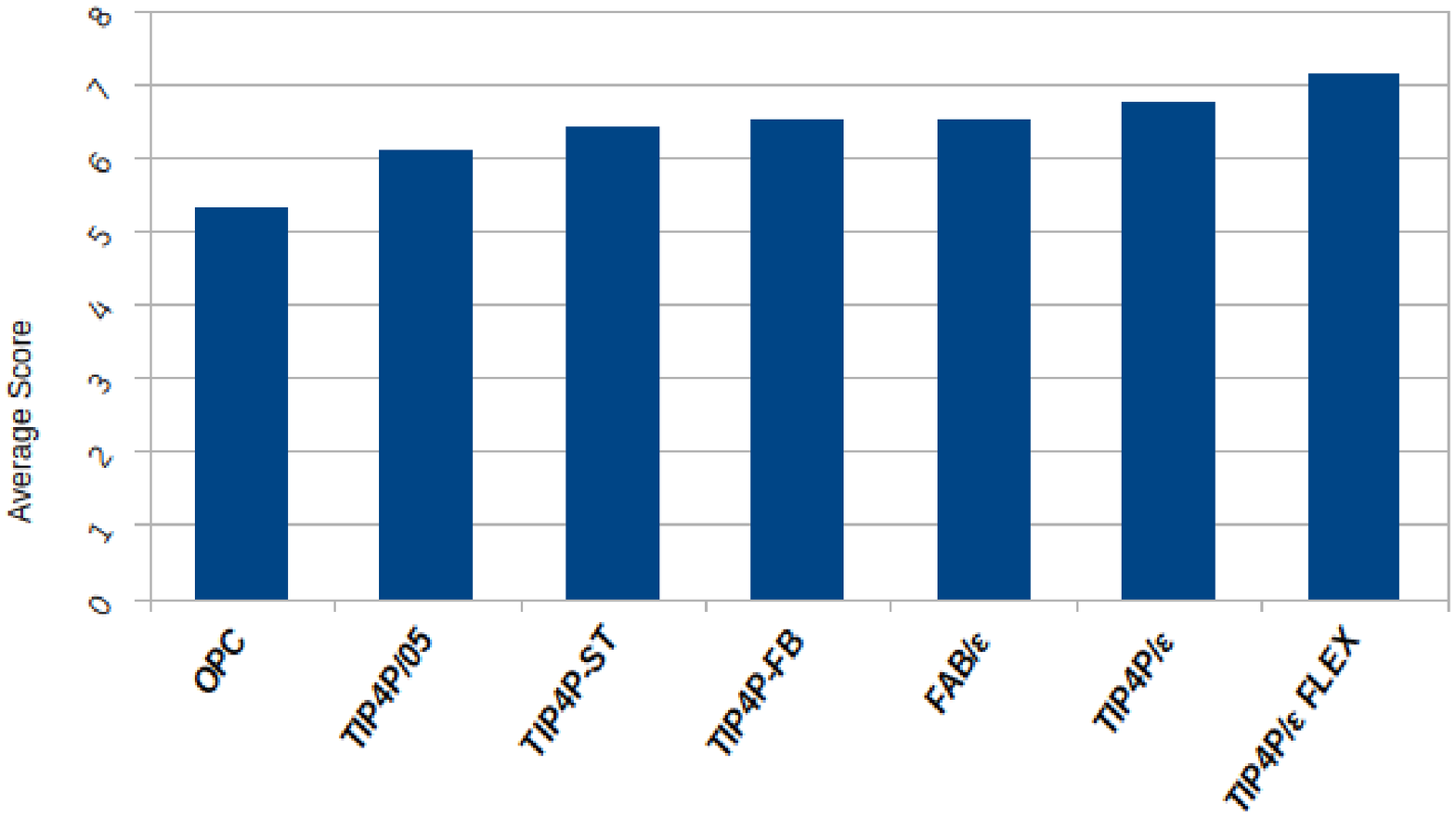,width=13.0cm,angle=0}}
\caption{Overall score (out of 10) }
\label{score}
\end{figure}
\newpage
The 4 sites rigid and non-polarizable models of water  have been widely studied and reparametrized since the TIP4P was published, achieving improvements of 53\% by the TIP4P/05, and reaching an optimization with the TIP4P/$\epsilon$ model and improving only 2.46\% with respect to TIP4P/05.
In the same timeline, there have been efforts to improve this type of model, such as the TIP4P-FB and OPC models, which as seen in figure \ref{score}, the TIP4P-FB improves the TIP4P/05 and is very close to the TIP4P/$\epsilon$ and the OPC that was optimized in the liquid phase, is far from the TIP4P/05 when it is analyzed in a wide and established range of properties in different thermodynamic phases.
By improving the 3 sites rigid and non-polarizable models with the FAB/$\epsilon$ model, a parametrization way is established with the IR spectrum as experimental data target and thus be able to add the harmonic potential and improve the reproduction of the experimental data.
As seen in table\ref{table3},\ref{table4} and in figure \ref{score}.\\
The TIP4P/05 model has been improved by FAB/$\epsilon$ an 7.1\% with respect to the properties that were already reproduced and the TIP4P/$\epsilon\hspace{0.1cm} _{Flex}$ improved an 15\% and open a new possibility by having flexibility and reproducing various properties with less error.
\begin{table}
\caption{•}
\label{table5}
\scalebox{0.8}[0.65]{
\begin{tabular}{|lccccccc|}
\hline\hline

 	\multicolumn{8}{|c|}{Average score (out of 10)}\\														
Property	&	TIP4P-FB	&	TIP4P-ST	&	OPC	&	TIP4P/05	&	TIP4P/$\epsilon$	&	FAB/$\epsilon$	&	TIP4P/$\epsilon\hspace{0.1cm} _{Flex}$	\\
\hline\hline
Enthalpy of phase change	&	8.2	&	8.2	&	5.5	&	6.7	&	7.9	&	5.9	&	7.2	\\
Critical point properties	&	5.4	&	5.1	&	4	&	7.6	&	4.9	&	3.5	&	9.3	\\
Surface tension	&	5.8	&	5.8	&	4.9	&	7.5	&	8.3	&	4.2	&	7.9	\\
Melting properties	&	7.1	&	7.3	&	1.8	&	8.3	&	7.4	&	7.7	&	6.7	\\
Orthobaric densities and TMD	&	9.5	&	9.8	&	9	&	8.9	&	9.6	&	9.3	&	9.5	\\
Isothermal compressibility	&	7.4	&	7.8	&	5.6	&	4.9	&	7.5	&	9	&	8.8	\\
Thermal expansion coefficient	&	8.9	&	8.3	&	8.6	&	9.8	&	9.1	&	6.7	&	8.7	\\
Heat capacity at constant pressure	&	4.3	&	4.5	&	5	&	3.3	&	4	&	5.2	&	0	\\
Liquid-vapor phase equilibrium (density)	&	5	&	4.9	&	4.4	&	6.2	&	4.8	&	5.3	&	7.4	\\
Gas properties	&	0	&	0	&	2.2	&	0	&	0	&	1.1	&	2.9	\\
Static dielectric constant	&	8.2	&	6.4	&	7.3	&	0	&	8.8	&	8.1	&	7.8	\\
T m -TMD-T c ratios	&	5.8	&	5.7	&	5.1	&	6.1	&	5.6	&	6.1	&	6.2	\\
Densities of ice Ih	&	7.1	&	7	&	2.8	&	7.5	&	7.4	&	8.1	&	6.7	\\
EOS high pressure	&	9.7	&	9.4	&	7	&	9.7	&	9.6	&	7.5	&	9.1	\\
Self-diffusion coefficient	&	8.9	&	8.2	&	8.7	&	7.6	&	7.7	&	5.8	&	8.1	\\
Shear viscosity	&	9.3	&	9	&	8.8	&	9.4	&	9.4	&	3.1	&	8.3	\\

\hline
\end{tabular}}
\end{table}
\newpage
The table \ref{table5} shows the average score of the properties studied with each model and it is notable that those of 4 sites rigid and non-polarizable models have more properties with a score lower than 5 and that the new TIP4P/$\epsilon\hspace{0.1cm} _{Flex}$ model only has two below 5, which are \textit{Heat capacity at constant pressure C$_P$ and Gas properties P$_{vap}$}, being one of the challenges for the next reparametrization of a water model or for any different type of model.

\newpage


\section{conclusions}
This work has important implications for the understanding of molecular
interactions in water and for the construction of molecular models in general, not only to implement the \textbf{$\mu_{min\rho}$} method and the IR spectrum as experimental target for harmonic potencial parametrization, but to understand the main idea of using the methodology to parameterize. By building a more closer to experimental data, but still approximate model, Understand which microscopic interactions are truly important for describing the properties of interest; for example, one part to highlight from this work is that the Dipole Moment of minimun density method  is capable  to construct force fields that define the optimal structure and capturing the effect of mutual polarization for a very wide range of water properties. Another interesting part is the role of model parameterization in establishing this understanding. For most practical problems it is impossible to explore the entire parameter space, so we can provide a way to reparameterization of water models and other molecules.

\newpage
{\bf Acknowledgements}\\
The author thank Consejo Nacional de Ciencia y Tecnolog\'ia (CONACYT) for financial support and 
 acknowledge Centro de Superc\'omputo of Universidad Aut\'onoma Metropolitana (Yoltla) for allocation of computer time. \\

\end{document}